\documentclass[prl,twocolumn,showpacs,preprintnumbers,amsmath,amssymb]{revtex4}
\usepackage{amsfonts}
\usepackage{bbm}
\usepackage{mathrsfs}
\usepackage{amsmath}
\usepackage[dvips]{graphics,color}
\usepackage{graphicx}
\usepackage{epstopdf}
\usepackage{dcolumn}
\usepackage{bm}
\usepackage{longtable}
\usepackage{graphics}
\usepackage{amssymb}
\usepackage{xspace}
\usepackage{epsfig}
\usepackage{subfigure}
\usepackage{dcolumn}
\makeatletter
\newcommand{\rmnum}[1]{\romannumeral #1}
\newcommand{\Rmnum}[1]{\expandafter\@slowromancap\romannumeral #1@}
\makeatother

\begin{document}
\title{Spontaneous Quantum Hall States in Chirally-stacked Few-Layer Graphene Systems}
\author{Fan Zhang}\email{zhangfan@physics.utexas.edu}
\author{Jeil Jung}
\author{Gregory A. Fiete}
\author{Qian Niu}
\author{Allan H. MacDonald}
\affiliation{Department of Physics, University of Texas at Austin, Austin TX 78712, USA}
\begin{abstract}
Chirally stacked $N$-layer graphene systems with $N \ge 2$ exhibit a variety of distinct broken symmetry states in which charge density contributions from
different spins and valleys are spontaneously transferred between layers.  We explain how these states are distinguished by their charge, spin,
and valley Hall conductivities, by their orbital magnetizations, and by their edge state properties.  We argue that valley Hall states have
$[N/2]$ edge channels per spin-valley.
\end{abstract}
\pacs{73.43.-f, 75.76.+j, 73.21.-b, 71.10.-w, 75.85.+t} \maketitle
\noindent {\em Introduction}--- In the early 1980s, following the discovery of the quantum Hall effect (QHE)\cite{QH}, it was
recognized\cite{TKNN} that electronic states can be characterized by topological indices, in particular by the integer valued Chern number
indices that distinguish quantum Hall states.  Quantum Hall states have non-zero Chern numbers and can occur only if time reversal symmetry
(TRS) is broken; until recently they have been observed only when TRS is explicitly broken by an external magnetic field. In this article we
discuss a class of broken symmetry states, first proposed theoretically\cite{Min_mf,Zhang_RG,Raghu_Hubbard_mf} and recently discovered
experimentally\cite{Yacoby_1,Yacoby_2}, which appear in chirally stacked graphene systems and are characterized by spin and valley dependent
spontaneous layer polarization.  The aim of the present paper is to explain how  these states are distinguished by their
charge\cite{QAH,QAH_RMP,Levitov}, spin\cite{Kane1}, and valley\cite{ValleyHall} quantized Hall conductances, by their orbital magnetizations,
and by their edge state properties.

Success in isolating monolayer and few-layer sheets from bulk graphite, combined with progress in the
epitaxial growth of few-layer samples, has opened up a rich new topic\cite{graphene_review1,graphene_review2} in two-dimensional electron
physics. Electron-electron interaction effects are most interesting in ABC-stacked $N \geq 2$ layer systems\cite{Mccann,Min_chiral,Zhang_ABC},
partly because\cite{Min_mf,Zhang_RG,Physics,Vafek,Levitov_2,jeil} their conduction and valence bands are very flat near the neutral system Fermi
level. For this special stacking order, low-energy electrons are concentrated on top and bottom layers and the low-energy physics of a $N$-layer
system is described approximately by a two band model with $\pm p^N$ dispersion and large associated momentum-space Berry
curvatures\cite{Berry}. When these band states are described in a pseudospin language, the broken symmetry state is characterized\cite{Min_mf}
by a momentum-space vortex with vorticity $N$ and a vortex-core which is polarized in the top-or-bottom layers. For AB stacked bilayers, for
example, interactions lead to a broken symmetry ground state\cite{Min_mf,Zhang_RG,Levitov_2} with a spontaneous gap in which charge is
transferred between top and bottom layers. ABC-stacked trilayer graphene has even flatter bands and is expected to be even more unstable to
interaction driven broken symmetries\cite{Zhang_ABC}, but samples that are clean enough to reveal its interaction physics have not yet been
studied.

\begin{figure}[htbp]
\centering{ \scalebox{0.50} {\includegraphics*{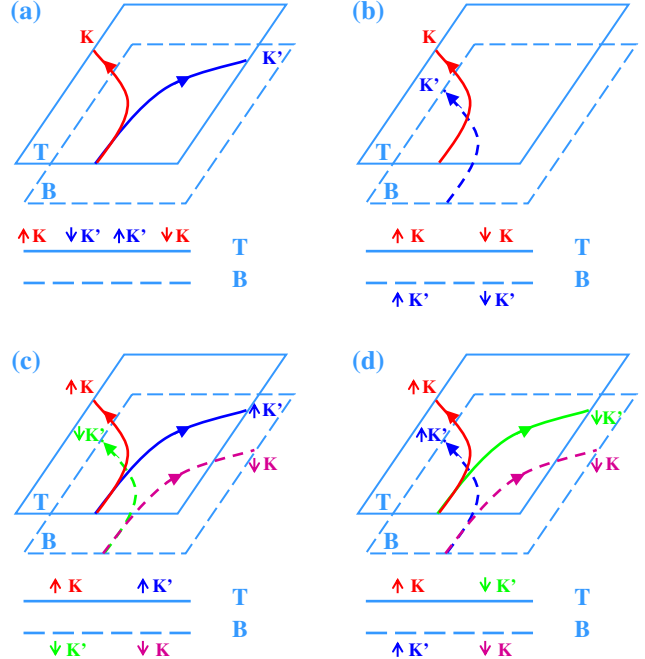}}}\caption{\label{fig:hall} {(Color online) For cases (a-d) the lower panel describes
the sense of layer polarization for each spin-valley combinations while the upper panel schematically indicates the corresponding Hall
conductivity contributions.
(a) a valley Hall insulator with a net layer polarization and a CDW mass $\lambda\sigma_{\rm z}$;
(b) an anomalous Hall insulator with a valley-dependent mass $\lambda\tau_{\rm z}\sigma_{\rm z}$;
(c) a layer-antiferromagnetic insulator with a spin-dependent mass $\lambda\,s_{\rm z}\sigma_{\rm z}$;
(d) a quantum spin Hall (or 2D Topological) insulator with a valley and spin
dependent mass term $\lambda\tau_{\rm z}s_{\rm z}\sigma_{\rm z}$.}}
\end{figure}
\begin{table}[h]
\caption{ Summary of spin-valley layer polarizations (T or B) and corresponding charge, spin, and valley Hall conductivities ($e^2/h$ units) and
insulator types for the three distinct states (b-d) with no overall layer polarization on which we focus, for a state in which every spin-valley
is polarized toward the top layer (a), and for a state with partial layer polarization. }
\newcommand\T{\rule{0pt}{3.1ex}}
\newcommand\B{\rule[-1.7ex]{0pt}{0pt}}
\begin{scriptsize}
\centering
\begin{tabular}{c | c c c c || c | c | c | c | c}
\hline\hline Fig. & $K\uparrow$ & $K\downarrow$ & $K'\uparrow$ & $K'\downarrow$
& $\sigma^{\rm (SH)}$ & $\sigma^{\rm (VH)}$ & $\sigma^{\rm (CH)}$ & $\sigma^{\rm (SVH)}$ & Insulator\T\\[3pt]
\hline
\ref{fig:hall}(b)  & T & T & B & B  &  $0$  &  $0$ & $2N$ & $0$ & QAH\T \\[3pt]
\ref{fig:hall}(c)  & T & B & T & B  &  $0$  &  $0$ & $0$ & $2N$ & LAF\T \\[3pt]
\ref{fig:hall}(d)  & T & B & B & T  &  $2N$ & $0$  &  $0$ & $0$ & QSH\T \\[3pt]
\ref{fig:hall}(a)  & T & T & T &  T  &  $0$ & $2N$ & $0$ & $0$ & QVH\T \\[3pt]
/  & T & T & T &  B  & $N$ & $N$ & $N$ & $N$ & All\T \\[3pt]
\hline\hline
\end{tabular}
\end{scriptsize}
\label{table:one}
\end{table}
\noindent {\em Hall Conductivities and Magnetizations}--- We discuss the electronic properties of $N$-layer ABC-stacked systems in terms of
ordered state mean-field Hamiltonians of the form,
\begin{eqnarray}
\label{eq:model} {\cal H}_{\rm N} =\frac{(v_0 p)^N}{(-\gamma_1)^{\rm N-1}}\big[\cos(N \phi_{\bm p})\sigma_{\rm x}+\sin(N \phi_{\bm p})
\sigma_{\rm y}\big] +\lambda\sigma_{\rm z}\,.
\end{eqnarray}
We have used the notation $\cos\phi_{\bm p}=\tau_z p_{\rm x}/p$ and $\sin\phi_{\bm p}=p_{\rm y}/p$ where $\tau_{\rm z}=\pm 1$ labels valleys $K$
and $K'$, the two inequivalent Brillouin zone corners. The Pauli matrices in Eq.~(\ref{eq:model}) act on a {\em which-layer} pseudospin
degree-of-freedom. In Eq.~(\ref{eq:model}) the first term\cite{Mccann,Zhang_ABC} is the low-energy $\bf{k} \cdot \bf{p}$ band Hamiltonian for a
single valley. Weak remote hopping processes have been dropped with the view that they do not play an essential role in the broken symmetry
states\cite{Zhang_RG}. The second term is an interaction-induced gap\cite{Min_mf,Zhang_RG,Levitov_2,jeil} term which defines the direction of
layer polarization in the momentum space vortex core.  For each spin and valley, symmetry is broken by choosing a sign for $\lambda$. We have
dropped the momentum dependence of $\lambda$ because, as we will see, it does not play an essential role below.  In the mean-field Hamiltonian
$2|\lambda|$ is the size of the spontaneous gap, $v_0$ is the Fermi velocity in monolayer graphene, and $\gamma_1\sim 0.4$ eV is the inter-layer
hopping energy. The $p^{\rm N}$ dispersion is a consequence of the $N$-step process in which electrons hop between low-energy sites in top and
bottom layers via high-energy states.

When spin and valley degrees-of-freedom are taken into account, the system has sixteen distinct broken symmetry states in which the sign of
$\lambda$ is chosen separately for $(K\uparrow )$, $(K\downarrow )$, $(K'\uparrow )$ and $(K'\downarrow )$ flavors. We take the view that any of
these states could be stable, depending on details that are beyond current knowledge and might be tunable experimentally. The sixteen states can
be classified according to their total layer-polarization which is proportional to the sum over spin-valley of the sign of $\lambda$. Six of the
sixteen states have no net layer charge transfer between top and bottom layers and are likely to be lowest in energy in the absence of an
external electric field. These six states can be separated into three doublets which differ only by layer inversion in every spin-valley. Thus
three essentially distinct states compete for the broken symmetry ground state: the anomalous Hall state in which the sign of $\lambda$ is
valley-dependent but not spin-dependent ($\lambda \sigma_{\rm z} \to \lambda \tau_{\rm z} \sigma_{\rm z}$), the layer-antiferromagnetic state in
which $\lambda$ is only spin-dependent ($\lambda \sigma_{\rm z} \to \lambda s_{\rm z} \sigma_{\rm z}$) and the topological insulator (TI) state
in which $\lambda$ is both spin and valley dependent ($\lambda \sigma_{\rm z} \to \lambda \tau_{\rm z} s_{\rm z}\sigma_{\rm z}$). These states
are distinguished by their spin and valley dependent Hall conductivities and orbital magnetizations indicated schematically in
Fig.~\ref{fig:hall} and summarized in Table~\ref{table:one}.

\begin{figure}[b]
\centering{ \scalebox{0.53} {\includegraphics*{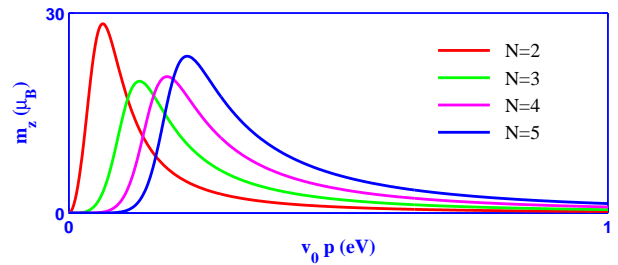}}}\caption{\label{fig:orb} {(Color online) The magnitude of orbital magnetic moments
carried by individual states versus in-plane momentum, for each spin and valley flavor in ABC graphene $N$-layers. Here the moments are in units
of $\mu_{\rm B}$ and $|\lambda|=10$ meV.}}
\end{figure}

\noindent {\em Berry Curvatures}--- The three broken symmetry states on which we focus are distinguished by the signs of the Berry
curvatures\cite{Berry} contributions from near the $K$ and $K'$ valleys of $\uparrow$ and $\downarrow$ spin bands; we note that the Berry
curvatures are non-zero only when inversion symmetry is spontaneously broken. Using the Berry curvatures, we evaluate the orbital magnetizations
and Hall conductivities of all three states. For momentum-independent mass $\lambda$ the Berry curvature of the $N$-layer chiral model is
\begin{eqnarray}\label{eq:berry}
\Omega^{(\rm \alpha)}_{\hat z}({\bm p},\tau_{\rm z},s_{\rm z})=-\alpha\frac{\tau_{\rm z}}{2}\frac{\lambda}{h_{\rm t}^3}\bigg(\frac{\partial
h_{\parallel}}{\partial p}\bigg)^2\,,
\end{eqnarray}
where symbol $\alpha=+(-)$ denotes the conduction (valence) band, and the transverse and total pseudospin fields are $h_{\parallel}=(v_0
p)^N/\gamma_1^{\rm N-1}$ and $h_{\rm t}=\sqrt{\lambda^2+h_{\parallel}^2}$. The orbital magnetic moment carried by a Bloch electron\cite{Berry}
is $m^{(\rm \alpha)}_{\hat{z}}=e\hbar\varepsilon^{(\rm \alpha)}\Omega^{(\rm \alpha)}_{\hat z}$ for a two-band model with particle-hole symmetry.
For the chiral band model
\begin{eqnarray}\label{eq:berry}
m^{(\rm \alpha)}_{\hat z}({\bm p},\tau_{\rm z},s_{\rm z})=\bigg[-\tau_{\rm z}\frac{\lambda}{h_{\rm t}^2}\bigg(\frac{\partial
h_{\parallel}}{\partial p}\bigg)^2 m_{\rm e}\bigg]\mu_{\rm B}\,,
\end{eqnarray}
where $m_{\rm e}$ is the electron mass and $\mu_{\rm B}$ is the Bohr magneton $e\hbar/2m_{\rm e}$.  Like the Berry curvature the orbital
magnetization changes sign when the valley label changes {\em and} when the sign of the mass term (the sense of layer polarization) changes,
{\em i.e.} both are proportional to $\tau_{\rm z} \rm{sgn}(\lambda)$. The orbital magnetization is however independent of the band index
$\alpha$. As illustrated in Fig.~\ref{fig:orb}, in the case of $|\lambda|=10$ meV, the orbital magnetic moment close to each Dirac point has a
symmetric sharp peak at which individual states carry moments twenty times larger than $\mu_{B}$.  The state in which $\lambda \to \lambda
\tau_{z}$ therefore has overall orbital magnetization and broken time reversal symmetry, even though it does not have a finite
spin-polarization.  Integrating over the valence band, we obtain a total orbital magnetization per area $\sim(N\lambda m_{\rm
e}/2\pi\hbar^2)\ln(\gamma_1/|\lambda|)\mu_{\rm B}$, that is $\sim 0.002 \mu_{\rm B}$ per carbon atom for $|\lambda|=10$ meV.

In the presence of an in-plane electric field, an electron acquires an anomalous transverse velocity proportional to the Berry curvature, giving
rise to an intrinsic Hall conductivity\cite{QAH_RMP,Berry}.  Using Eq.~(\ref{eq:berry}), we find that the intrinsic Hall conductivity
contribution from a given valley and spin is
\begin{eqnarray}\label{eq:hall}
\sigma_{\rm H}^{(\rm \alpha)}(\tau_{\rm z},s_{\rm z})=\frac{\tau_{\rm
z}}{2}\frac{Ne^2}{h}\bigg(\frac{\lambda}{h_{t}\left( p_F \right)}-\frac{\lambda}{|\lambda|}\delta_{\rm \alpha,+}\bigg)\,,
\end{eqnarray}
where $h_t (p_F)$ is the total pseudospin field at the Fermi wavevector.
Provided that the Fermi level lies in the mass gap, each spin and valley
contributes $Ne^2/2h$ to the Hall conductivity, with the sign given by $\tau_{\rm z} \rm{sgn}(\lambda)$.

In Fig.~\ref{fig:hall}(a) we consider the case in which each spin-valley is polarized in the same sense.  The total Hall conductivity is then
zero for both spins, with the K and K' valleys making Hall conductivity and magnetization contributions of opposite sign, preserving time
reversal symmetry. This phase can be viewed as having a valley Hall effect\cite{ValleyHall} and, even though it does not break time-reversal
symmetry, we argue later that this designation has physical significance.

As shown in Fig.~\ref{fig:hall}(b), the case $\lambda \sigma_{\rm z} \to \lambda \tau_{\rm z}\sigma_{\rm z}$ implies Hall conductivity and
orbital magnetization contributions of the same sign for each spin and valley.  This state breaks time reversal symmetry but its spin density is
surprisingly is everywhere zero.  The total Hall conductivity has the quantized value $2Ne^2/h$. Similarly, the orbital magnetic moment has the
same sign for all flavors.  We refer to this state as the quantized anomalous Hall state.  In addition to its anomalous Hall effect, this state
has a substantial orbital magnetization. The anomalous Hall states is probably most simply identified experimentally by observing a $\nu=2N$ QHE
which persists to zero magnetic field.

For $\lambda \sigma_{\rm z} \to \lambda s_{\rm z} \sigma_{\rm z}$, depicted in Fig.~\ref{fig:hall}(c) the two spins have valley Hall effects of
opposite sign, and the two layers have spin-polarizations of opposite sign. This layer-antiferromagnetic state has broken time reversal
symmetry and opposite spin-polarizations on top and bottom layers.

Fig.~\ref{fig:hall}(d) describes the third type of state with effective interaction $\lambda \sigma_{\rm z} \to \lambda \tau_{\rm z} s_{\rm
z}\sigma_{\rm z}$. This state does not break time reversal invariance and instead has anomalous Hall effects of opposite signs in the two spin
subspaces, i.e. a spin Hall effect.  Neither the top nor the bottom layer has spin or valley polarization. Quite interestingly if we only
consider one layer, there are both spin Hall and valley Hall effects, however, the orientations of the Hall currents in the top and the bottom
layers are the same for the spin Hall effects but opposite for the valley Hall effects.

Table~\ref{table:one} includes as well the case in which one flavors polarizes in the opposite sense of the other three; charge, valley, and
spin Hall effects coexist in this state which can be favored by a small potential difference between the layers.

\begin{figure}[htbp]
\centering \scalebox{0.55} {\includegraphics* [0.75in,3.50in][6.70in,8.70in] {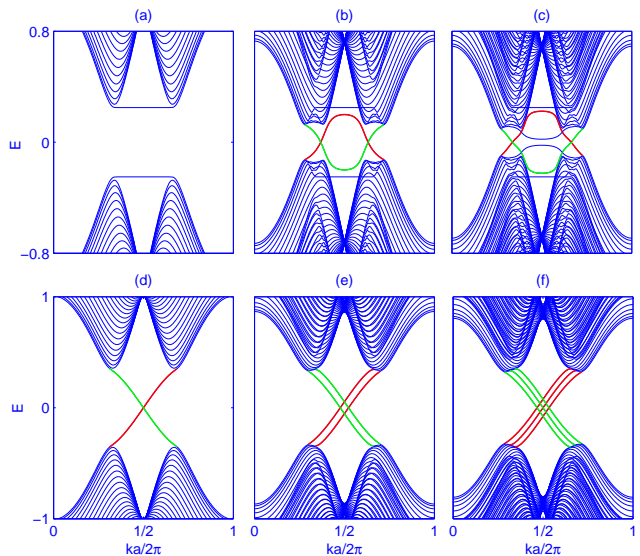}}\caption{\label{fig:edgestates} {(Color online)
Intra-valley and inter-valley edge states in chirally-stacked graphene systems. (a)(d) for a single layer, (b)(e) for a bilayer and (c)(f) for a
trilayer. To visualize the edge states, the intralayer and interlayer nearest neighbor hoppings are chosen as $\gamma_0=1$ and $\gamma_1=0.3$,
respectively; $\lambda=0.25$ for Fig.(a,b,c) and $\lambda=0.3\sqrt{3}\,\tau_{\rm z}$ for Fig.(d,e,f).}}
\end{figure}
\noindent{\em Edge States}--- The physical significance of spontaneous charge, valley, and spin anomalous Hall effects is illustrated in
Fig.~\ref{fig:edgestates}. Graphene has very weak spin-orbit interactions, which in our case we ignore altogether.  Fig.~\ref{fig:edgestates}
compares the edge electronic structure of $N=1,2,3$ spinless models with a quantized anomalous Hall effect ({\em i.e.} with opposite layer
polarizations at two valleys) and with a quantized valley Hall effect. The states with anomalous Hall effects have $N$ topologically protected
robust chiral edge states associated with the QHE, as shown in Fig.~\ref{fig:edgestates}(d)(e)(f). The edge state structure associated with the
valley Hall states is more interesting.  In the $N=1$ valley Hall state the Hall conductivity contribution associated with each valley is $1/2$
in $e^2/h$ units; the full unit of Hall conductance requires the two valleys to act in concert.  Because they act in opposition in the valley
Hall state, there is no edge state, as shown in Fig.~\ref{fig:edgestates}(a). For $N=2$ on the other hand, each valley contributes a full
quantum Hall effect, and as we see in Fig.~\ref{fig:edgestates}(b) we find two chiral edge states with opposite chirality, one associated with
each valley. For $N=3$ depicted in Fig.~\ref{fig:edgestates}(c), the additional half quantum Hall effect from each valley is insufficient to
produce a new chiral edge state branch. In general we expect $[N/2]$ chiral edge state branches at each valley in an $N$-layer stack. Of course
valley Hall edge states are topologically protected only when the edge-direction projections of $K$ and $K'$ valleys are not coincident and
inter-valley scattering due to disorder is absent. Nevertheless, we expect robust edge states to be present in valley Hall states, as
found\cite{Morpurgo} in numerical studies of valley Hall states induced by an external electric field without interactions.

\noindent {\em Discussion}--- At the level of continuum-model mean-field theory\cite{Min_mf}, the three charge balanced states we have discussed
are degenerate. In addition to breaking inversion symmetry, each breaks two of three additional symmetries; time reversal ($\mathcal{T}$), spin
rotational invariance ($SU(2)$), and the valley Ising symmetry ($\mathcal{Z}_2$). The TI state preserves only $\mathcal{T}$, the AH phase
preserves only spin-rotational invariance, and the AF state has $\mathcal{Z}_2$ symmetry. Both TI and AF phases break the continuous $SU(2)$
symmetry and therefore Goldstone modes emerge\cite{Levitov}. The actual ground state is dependent on subtle correlation and microscopic physics
issues that are beyond the scope of this paper. We note however that it might be possible to induce transitions between different possible
states using external fields. For example, the energy of the quantized anomalous Hall state will be lowered by a perpendicular external magnetic
field because it has a large orbital magnetization. The fully layer polarized state will be favored by an external electric field which produces
a potential difference between the layers. Increasing the magnetic field further results in quantum Hall
ferromagnetism\cite{QH_Yafis,QH_Yacoby,QH_Kim}. Recent experiments\cite{Yacoby_1,Yacoby_2} in bilayers appear to provide definitive proof that
the ground state at very weak external magnetic fields is the quantized anomalous Hall state.

The quantum spin Hall effect we discuss in this paper is in several respects different from that discussed in the well known
papers\cite{Kane1,Raghu_Hubbard_mf} which foreshadowed the identification of topological insulators. (\rmnum{1}) The quantum spin Hall effect is
driven by broken symmetries produced by electron-electron interactions, rather than by spin-orbit interactions\cite{Kane1} which we neglect. The
effective spin-orbit coupling $\lambda\tau_{\rm z}s_{\rm z}\sigma_{\rm z}$ due to electron-electron interactions can be $10^4$ times larger than
the intrinsic one\cite{Hongki_so}. (\rmnum{2}) Unlike the previous interaction induced TI phase\cite{Raghu_Hubbard_mf} which appears only at
finite interaction strengths, here the instability to the TI phase is present even for weak interactions. (\rmnum{3}) The broken symmetry occurs
only for $N \ge 2$, rather than in the single-layer systems\cite{Kane1,Raghu_Hubbard_mf}. (\rmnum{4}) Our states are also distinguished
topologically, since they are characterized by Chern numbers which can have any integer value, rather than by a $Z_2$ label. Of course, only
N-odd layers are strong TI's, because the helical edge modes are likely to localize in a $N$-even system due to the backscattering process
allowed by $\mathcal{T}$\cite{Xu}.  Even though spin-orbit interactions are very weak in graphene\cite{Hongki_so}, our more elaborate Chern
number state classification is therefore not strictly valid once they are included. In this case the quantum spin Hall effect will no longer be
precisely quantized even for odd $N$.

\noindent{\em Acknowledgements}--- This work has been supported by Welch Foundation under grant TBF1473, NRI-SWAN, DOE grant Division of
Materials Sciences and Engineering DE-FG03-02ER45958, NSF under grant DMR-0606489 and ARO W911NF-09-1-0527. We acknowledge helpful discussions
with C.L. Kane and J. Wen.


\begin{thebibliography}{100}
\bibitem{QH}
 K. Von Klitzing, G. Dorda, and M. Pepper, {Phys. Rev. Lett.} {\bf 45}, 494 (1980).

\bibitem{TKNN}
 D. J. Thouless, M. Kohmoto, M. P. Nightingale, and M. Dennijs, Phys. Rev. Lett. {\bf 49}, 405 (1982).

\bibitem{Min_mf}
 H. Min, G. Borghi, M. Polini and A. H. MacDonald, {Phys. Rev. B} {\bf 77}, 041407(R) (2008).

\bibitem{Zhang_RG}
 F. Zhang, H. Min, M. Polini and A. H. MacDonald, {Phys. Rev. B} {\bf 81}, 041402(R) (2010).

\bibitem{Raghu_Hubbard_mf} Similar broken symmetry states in hexagonal lattice
Hubbard models were proposed independently by S. Raghu, X. Qi, C. Honerkamp, and S. Zhang, {Phys. Rev. Lett.} {\bf 100}, 156401 (2008); J. Wen,
A. Ruegg, C. Wang, and G. A. Fiete, Phys. Rev. B {\bf 82}, 075125 (2010).

\bibitem{Yacoby_1}
 J. Martin {\it et al.}, arXiv:1009.2069 (2010).

\bibitem{Yacoby_2}
 R. T. Weitz {\it et al.}, arXiv:1010.0989 (2010).

\bibitem{QAH}
 F. D. M. Haldane, {Phys. Rev. Lett.} {\bf 61}, 2015 (1988).

\bibitem{QAH_RMP}
 N. Nagaosa, J. Sinova, S. Onoda, A. H. MacDonald and N. P. Ong, {Rev. Mod. Phys.} {\bf 82}, 1539 (2010).

\bibitem{Levitov}
 R. Nandkishore and L. Levitov, Phys. Rev. B {\bf 82}, 115124 (2010).

\bibitem{Kane1}  The quantum spin Hall effect in these systems should be distinguished
from the quantum spin Hall effect discussed in the context of two-dimensional topological insulators,
for example in C. L. Kane and E. J. Mele, {Phys. Rev. Lett.} {\bf 95}, 226801 (2005).
Topological insulators do not have broken TRS, and do not normally exhibit a
quantized bulk spin Hall effect, but do have topologically protected edge states similar to those of
quantum Hall systems.  The edge states of two-dimenstional topological insulators have been
studied experimentally.  See M. Konig {\it et al.}, Science {\bf 318}, 766 (2007).

\bibitem{ValleyHall}
 D. Xiao, W. Yao and Q. Niu, {Phys. Rev. Lett.} {\bf 99}, 236809 (2007).

\bibitem{graphene_review1}
 A. K. Geim and A. H. MacDonald, {Phys. Today} {\bf 60}(8), 35(2007).

\bibitem{graphene_review2}
 A. H. Castro Neto {\it et al.}, {Rev. Mod. Phys.} {\bf 81}, 109 (2009).

\bibitem{Mccann}
 E. McCann and V. I. Fal'ko, {Phys. Rev. Lett.} {\bf 96}, 086805 (2006).

\bibitem{Min_chiral}
 Hongki Min and A. H. MacDonald, {Phys. Rev. B} {\bf 77}, 155416 (2008).

\bibitem{Zhang_ABC}
 F. Zhang, B. Sahu, H. Min, A. H. MacDonald,  {Phys. Rev. B} {\bf 82}, 035409 (2010).

\bibitem{Physics}
 F. Guinea, Physics {\bf 3}, 1 (2010).

\bibitem{Vafek}
 O. Vafek and K. Yang, {Phys. Rev. B} {\bf 81}, 041401(R) (2010).

\bibitem{Levitov_2}
 R. Nandkishore and L. Levitov, {Phys. Rev. Lett.} {\bf 104}, 156803 (2010).

\bibitem{jeil}
 J. Jung, F. Zhang, A. H. MacDonald, arXiv:1010.1819 (2010).

\bibitem{Berry}
 D. Xiao, M. Chang and Q. Niu, {Rev. Mod. Phys.} {\bf 82}, 1959 (2010).

\bibitem{Morpurgo}
 J. Li, I. Martin, M. Buttiker, A. F. Morpurgo, arXiv:1009.4851 (2010).

\bibitem{QH_Yafis}
 Y. Barlas, R. Cote, K. Nomura and A. H. MacDonald, Phys. Rev. Lett. {\bf 101}, 097601(2008).

\bibitem{QH_Yacoby}
 B. Feldman, J. Martin and A. Yacoby, Nature Phys. {\bf 5}, 889 (2009).

\bibitem{QH_Kim}
 Y. Zhao, P. Cadden-Zimansky, Z. Jiang and P. Kim, Phys. Rev. Lett. {\bf 104}, 066801 (2010).

\bibitem{Hongki_so}
 H. Min {\it et al.}, {Phys. Rev. B} {\bf 74}, 165310 (2006).

\bibitem{Xu}
 C. Xu and J. E. Moore, {Phys. Rev. B} {\bf 73}, 064417 (2006); E. Prada, P. San-Jose and L. Brey, arXiv:1007.4910 (2010).
\end{thebibliography}
\end{document}